\documentclass{eptcs}
\providecommand{\event}{REFINE 2015} 

\usepackage[usenames,dvipsnames]{color}
\usepackage[color]{czt}
\usepackage[color]{circus}
\usepackage{amssymb}
\usepackage{listings}
\usepackage{graphicx}
\usepackage{stmaryrd}
\usepackage{url}

\usepackage[T1]{fontenc}
\newcommand{\meta}[1]{\textrm{\guillemotleft} #1 \textrm{\guillemotright}}

\newcommand{\CircusTime}{\textsf{\slshape Circus Time}}
\newcommand{\OhCircus}{\textsf{\slshape OhCircus}}
\newcommand{\SCJCircus}{\textsf{\slshape SCJ-Circus}}

\newenvironment{syntax}{\begin{displaymath}\begin{array}{lcl}}{\end{array}\end{displaymath}}
\newcommand{\sdef}{& \quad \mathsf{::=} \quad &}

\newcommand{\scat}[1]{\mathsf{#1}}

\renewcommand{\circinterrupt}{\mathbin{\triangle}}

\newcommand{\circstartby}{\mathbin{\blacktriangleleft}}
\newcommand{\circendby}{\mathbin{\blacktriangleright}}

\newcommand{\circinitial}{\mathbf{initial}}

\newcommand{\circnew}{\mathbf{new}}

\newcommand{\circstart}{\mathbf{start}}
\newcommand{\circperiod}{\mathbf{period}}
\newcommand{\circsafelet}{\mathbf{safelet}}
\newcommand{\circsequencer}{\mathbf{sequencer}}
\newcommand{\circmission}{\mathbf{mission}}
\newcommand{\circhandler}{\mathbf{handler}}
\newcommand{\circperiodic}{\mathbf{periodic~handler}}
\newcommand{\circaperiodic}{\mathbf{aperiodic~handler}}
\newcommand{\circgetSequencer}{\mathbf{getSequencer}}

\newcommand{\circinitialize}{\mathbf{initialize}}

\newcommand{\circhandleAsyncEvent}{\mathbf{handleAsyncEvent}}

\newcommand{\circreturn}{\mathbf{return}}
\newcommand{\circsequencerref}{\mathbf{sequencer}}

\newcommand{\circnewI}{\mathbf{newI}}
\newcommand{\circnewM}{\mathbf{newM}}
\newcommand{\circnewPR}{\mathbf{newPR}}
\newcommand{\circnewPM}{\mathbf{newPM}}

\renewcommand{\circblockbegin}{\left(\begin{array}{l}}
\renewcommand{\circblockend}{\end{array}\right)}

\renewcommand{\circspot}{@}
\renewcommand{\circseq}{;}
\renewcommand{\circthen}{\longrightarrow}
\renewcommand{\circmu}{\mu}
\renewcommand{\Semi}{\textrm{\large ;}~}

\title{\SCJCircus: a refinement-oriented formal notation for Safety-Critical Java}
\author{Alvaro Miyazawa \qquad\qquad Ana Cavalcanti
	\institute{Department of Computer Science, University of York, York, YO10 5GH, UK}
	\email{alvaro.miyazawa@york.ac.uk \qquad\qquad ana.cavalcanti@york.ac.uk}
}

\def\titlerunning{\SCJCircus}
\def\authorrunning{A. Miyazawa \& A. Cavalcanti}

\begin{document}

\maketitle

\begin{abstract}
Safety-Critical Java~(SCJ) is a version of Java whose goal is to support the development of real-time, embedded, safety-critical software. In particular, SCJ supports certification of such software by introducing abstractions that enforce a simpler architecture, and simpler concurrency and memory models. In this paper, we present \SCJCircus, a refinement-oriented formal notation that supports the specification and verification of low-level programming models that include the new abstractions introduced by SCJ. \SCJCircus~is part of the family of state-rich process algebra \Circus, as such, \SCJCircus\ includes the \Circus\ constructs for modelling sequential and concurrent behaviour, real-time and object orientation. We present here the syntax and semantics of \SCJCircus, which is defined by mapping \SCJCircus\ constructs to those of standard \Circus. This is based on an existing approach for modelling SCJ programs. We also extend an existing \Circus-based refinement strategy that targets SCJ programs to account for the generation of \SCJCircus~models close to implementations in SCJ.
\end{abstract}

\section{Introduction}

Safety-Critical Java (SCJ)~\cite{SCJ} is a subset of the Real-Time Specification for Java (RTSJ) \cite{RTSJ}. This is a version of Java that targets the development of real-time software. It avoids the issue of unpredictable timing associated with garbage collection by introducing memory areas.

SCJ restricts the RTSJ to facilitate certification; it imposes a particular structure for programs embedding simplified memory and concurrency models. The structure of an SCJ application is composed of a safelet (the main program), a mission sequencer that provides missions in a particular order, and a number of missions that are composed by concurrent handlers. SCJ supports different types of handlers, such as periodic and aperiodic handlers.

\begin{figure}[!htb]\centering
\includegraphics[width=0.8\textwidth]{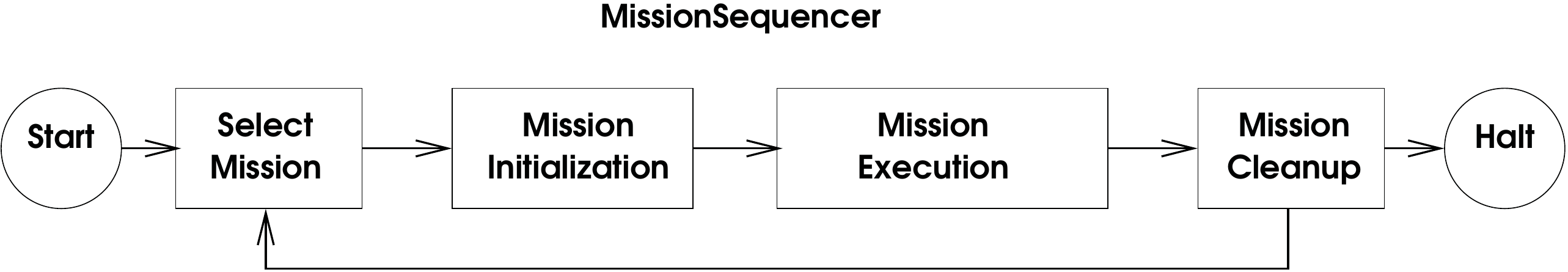}
\caption{SCJ programming model}
\label{fig:programming-model}
\end{figure}

\noindent Figure~\ref{fig:programming-model} depicts the programming model of SCJ, which essentially consists of a cycle where at each step a new mission is selected, initialised, executed and terminated. During execution, a number of handlers run in parallel. When there are no missions left, the program terminates. The memory model is based on scoped memory regions, rather than garbage collection.  The safelet, the missions, and the handlers have associated memory regions, which are cleared at predictable points of the program control flow.  

In \cite{Cavalcanti2013}, the SCJ standard is complemented with a design technique based on the \Circus~family of languages for refinement. \Circus~\cite{Circus} is a state-rich process algebra for refinement that has been applied to the verification of a variety of models including Simulink and Stateflow diagrams~\cite{CCO05,MC12}. The semantics of \Circus~is based on Hoare and He's Unifying Theories of Programming~\cite{UTP}, which is a semantic framework that supports the formalisation of paradigms in an independent fashion and their combination through specific techniques. As refinement is a central concept in UTP, it is also an important aspect of \Circus~as evidenced by its rich refinement calculus~\cite{Oli06}. \Circus~has been extended to support a number of different programming paradigms: for example, \OhCircus~\cite{OhCircus} supports the specification of object-oriented programs, and \CircusTime~\cite{WWC12} supports modelling real-time programs.

We introduce here a new member of the set of \Circus\ languages: \SCJCircus~combines \OhCircus~and \CircusTime, and extends them with the abstractions introduced by SCJ. It supports either verification or full development of SCJ programs from an abstract timed-model to an object-oriented timed model that explicitly uses the SCJ abstractions. \SCJCircus\ models define a safelet, a mission sequencer, missions and handlers. Additionally, \SCJCircus~introduces object creation statements ($\circnew$ in \OhCircus) tailored to the hierarchical memory model adopted in Safety-Critical Java. Abstraction can still be achieved using the constructs of \Circus\ for data and behavioural modelling. Yet, the architecture of the models is in direct correspondence with that of SCJ programs, although platform specific aspects of an application, such as memory and thread availability, are not covered.

The refinement strategy proposed in \cite{Cavalcanti2013} is based on the notion of anchors, which are models written in different subsets of \Circus\ following specific architectural patterns. There are four anchors related by refinement: A, O, E and S. The first anchor~(A anchor) defines an abstract model and the last anchor~(S anchor) describes a refinement of the A anchor that follows the programming paradigm of SCJ. The O anchor introduces the object-oriented model, and the E anchor introduces the notions of missions, handlers and memory areas. Whilst \cite{Cavalcanti2013} details the refinement strategy between the three first anchors~(A, O and E), it only briefly indicates how to proceed from the E to the S anchor.

In this work, we extend \cite{Cavalcanti2013} exploring the use of \SCJCircus\ to define the S anchor. We specify the syntax and semantics of \SCJCircus, and describe the last phase of the refinement strategy to use \SCJCircus\ as target models. To define the semantics of \SCJCircus, we build on a \Circus\ semantics of SCJ programs defined in~\cite{ZLCW13}. To that end, we update that semantics to reflect fundamental changes to the mode of interaction between handlers and mission termination. We also propose a different structure for the \Circus\ models to enable compositional refinement with respect to the \SCJCircus\ components.

In Section~\ref{sec:preliminary}, we introduce the \Circus~family of languages and Safety-Critical Java. In Section~\ref{sec:scjcircus}, we discuss \SCJCircus, its syntax and semantics, and Section~\ref{sec:refinement} discusses the extension of the refinement strategy proposed in \cite{Cavalcanti2013} to reach \SCJCircus~programs. Finally, Section~\ref{sec:conclusions} concludes by relating our work to the existing literature and discussing future work.

\section{Preliminaries}
\label{sec:preliminary}

In this section, we briefly describe the base notations relevant to our work. Section~\ref{subsec:circus} introduces the \Circus\ family of languages, and Section~\ref{subsec:scj} describes SCJ.

\subsection{\Circus}
\label{subsec:circus}

In this section, we use the \CircusTime~process $PEHFW$~(periodic event handler framework) in Figure~\ref{fig:PEHFW} that models the general behaviour of a periodic event handler to describe \Circus\ and its timed variant. The main modelling element of a \Circus\ specification is a process (indicated by the keyword $\circprocess$) that declares state components (identified by the keyword $\circstate$), a number of auxiliary actions, and a main action (prefixed by $\circspot$) that describes the overall behaviour of the process. In the case of our example, the process $PEHFW$ is parametrised by an identifier $id$ of a given type $ID$, and declares two state components, $start$ and $period$ both of type $\nat$. 

\begin{figure}
	\hspace{-1cm}\begin{minipage}{\textwidth}
		\begin{circus}
			\circprocess PEHFW \circdef id: ID \circspot \circbegin\\\quad
			\circstate PEHFWState == [start,period:\nat]\\\quad
			Execute \circdef \circwait start\circseq\\\quad
			\circblockbegin
			\circmu X @ \circblockbegin
			(handleAsyncEventCall!id!id \then handleAsyncEventRet!id!id \then Skip) \circendby period \circseq X\\
			\extchoice\\
			done\_handler!id \then \Skip
			\circblockend\\
			\lpar \{\} | \lchanset handleAsyncEventCall.id, done\_handler.id \rchanset | \{\} \rpar\\
			(\circmu Y @ ((handleAsyncEventCall!id!id \then \circwait period) \circstartby 0)\circseq Y) \circinterrupt done\_handler!id \then \Skip
			\circblockend\\\t1
			\circspot \circmu X @ (start\_peh?o!id?s?p \then 
			(start := s \circseq period := p) \circseq Execute\circseq X)\\
			\circend\\
		\end{circus}
	\end{minipage}\vspace{-.5cm}
	\caption{Framework process of the periodic event handler.}
	\label{fig:PEHFW}
\end{figure}

$PEHFW$ only declares one auxiliary action $Execute$. Actions are specified using a combination of Z~\cite{WD96} for data modelling and CSP~\cite{Ros11} for behavioural descriptions. The main action is defined by a recursion ($\circmu X @ \ldots$) that at each step starts an instance of the event handler via a communication through channel $start\_peh$. In this communication, the identifier $o$ of the mission that requested the instantiation is input, the handler identifier $id$ is output, and its start time $s$ and period $p$ are input. Whilst the input $o$ is not needed for the execution, it is necessary to allow missions to reuse periodic event handlers in the same application. The values of $s$ and $p$ are then assigned to the state components $start$ and $period$. The execution of a newly created handler is defined by the action $Execute$. It first waits for $start$ time units~($\circwait~start$), and then starts two recursive actions in parallel~($A_1\lpar ns_1|cs|ns_2\rpar A_2$) synchronising on channels $handleAsyncEventCall$ and $done\_handler$.

The first action specifies that at each step of the recursion there is an external choice~($\extchoice$) for communication on channels $done\_handler$ or $handleAsyncEventCall$ in this way, the handler can be terminated with a choice of $done\_handler$ or the method \texttt{handleAsyncEvent} is called through the channel $handleAsyncEventCall$. If \texttt{handleAsyncEvent} is called, the it must return, as indicated with a communication via $handleAsyncEventRet$ within $period$ time units. This is specified by the \CircusTime\ action $A\circendby e$ that defines that the action $A$ must terminate within $e$ time units. 

The recursive parallel action in $Execute$ adds a requirement that a call to \texttt{handleAsyncEvent} must be started as soon as it is available, and should only be made available again after $period$ time units. This is achieved by imposing a restriction on the communication $handleAsyncEventCall$ using the start by operator ($\circstartby$) that specifies that an action must start within a certain number of time units. The requirement that $handleAsyncEventCall$ is only offered after $period$ time units is enforced by the action $\circwait period$ after the communication on $handleAsyncEventCall$. Since the first recursion can be terminated by a synchronisation on the channel $done\_handler$, the second recursion must also be terminated. This is achieved by allowing its interruption of the recursion by a synchronisation on $done\_handler$ using the interrupt operator~($\circinterrupt$).

In general, a \Circus\ or \CircusTime\ specification consists of a sequence of paragraphs that define processes (as well as channels, constants, and other constructs that support the definition of processes). Processes are used to define the system and its components: state is encapsulated and interaction is via channels. Processes can be composed, via CSP operators, to define other processes. In \CircusTime, wait and deadline operators can be used to define time restrictions. In \OhCircus\ models, we can in addition define paragraphs that declare classes used to define types. More information about these languages can be found in \cite{Oli06,WWC12,OhCircus}. In the sequel, we further explain the notation as needed.

\subsection{Safety Critical Java}
\label{subsec:scj}

As previously mentioned, an SCJ application is formed by a safelet, mission sequencer, a number of missions, and periodic and aperiodic event handlers. Each of these is characterised by an interface or abstract class of an API that supports the development of SCJ programs via implementation and extension of these components. A safelet instantiates a mission sequencer, and iteratively obtains a mission from the mission sequencer, executes it and waits for it to terminate. The execution of a mission consists of the parallel execution of all its periodic and aperiodic event handlers. Most of the actual behaviour of the application is concentrated in the handlers, which are the focus of this section.

Our running example is a simple SCJ application: a communication medium that checks whether the three copies of a message received are the same (and, therefore, reliable). It has a single mission containing two handlers: one periodic event handler and one aperiodic event handler. The periodic event handler reads an input every at every cycle, stores it in a buffer, and releases the aperiodic event handler. Upon release, the aperiodic event handler examines the last three elements and outputs ``true'' or ``false'' depending on whether the last three values of the buffer are all the same or not.

\begin{figure}[!t]
\begin{lstlisting}[basicstyle=\small]
	public class Checker extends AperiodicEventHandler {
	  Buffer buffer;
	  public Checker(Buffer b) {
	    super(new PriorityParameters(Priorities.PR98),
	          new AperiodicParameters(),
	          storageParameters_Handlers);
	    buffer = b;
	  }
	  public void handleAsyncEvent() {
	    if (buffer.theSame()) devices.Console.println("true");
	    else devices.Console.println("false");
	  }
	}
\end{lstlisting}\vspace{-.5cm}
\caption{SCJ Level 1 example: Aperiodic Event Handler}
\label{fig:example-aeh}
\end{figure}

Figure~\ref{fig:example-aeh} shows the code for the aperiodic handler in our example. It extends the SCJ API class \texttt{AperiodicEventHandler}, and declares a local variable \texttt{buffer}, a constructor that receives an instance of the class \texttt{Buffer} and assigns it to \texttt{buffer}, and a \texttt{handleAsyncEvent} method that defines the main behaviour of the handler. The constructor of \texttt{Checker} calls the constructor of the superclass with priority~98, a new aperiodic parameter, and storage parameters that specify the amount of memory used by the handler. The method \texttt{handleAsyncEvent} checks whether the last three elements of \texttt{buffer} are the same using the method \texttt{theSame}; if they are, it prints ``true'', otherwise it prints ``false''. For simplicity, we print the output of the checker, which in practice needs to be sent to another component of the system. 

The complete program contains classes to implement the safelet, the mission sequencer, the mission and the periodic handler. It can be found in~\url{http://www.cs.york.ac.uk/~alvarohm/er2015.zip}.

\section{\SCJCircus}
\label{sec:scjcircus}

As previously mentioned, \SCJCircus~extends \OhCircus~and \CircusTime~with abstractions that are specific to Safety-Critical Java. Below, Section~\ref{sec:syntax} briefly discusses the syntax of \SCJCircus, Section~\ref{sec:model} presents the semantic models of the SCJ framework, that is, its API and programming model, and Section~\ref{sec:semantics} describes the semantics of the language based on the \Circus~models of Section~\ref{sec:model}.

\subsection{Syntax}
\label{sec:syntax}
\SCJCircus\ extends the syntax of \OhCircus~and \CircusTime~with paragraphs that allow the specification of safelets, mission sequencers, missions and handlers. Figure~\ref{fig:S-anchor} presents the specification of our running example in \SCJCircus. It matches the structure of our example, but further specifies timing requirements. The periodic event handler reads an input every $P$ time units, with an input deadline of $ID$ time units. Each cycle of the periodic event handler takes any time between $0$ and $PTB$ time units, and must terminate within $PD$ time units. The aperiodic event handler outputs values within $OD$ time units, and each release takes at most $ATB$ time units, and must terminate within $AD$ time units.

The constants $PTB$, $ATB$, $ID$, $OD$, $PD$, $AD$ and $P$ need to satisfy a number of conditions to ensure that the two handlers run in lockstep. For the periodic event handler, these conditions require that the sum of periodic time budget ($PTB$) and the input deadline ($ID$) does not exceed the periodic deadline ($PD$). Additionally the sum of the periodic deadline ($PD$) and the aperiodic deadline $AD$ must not exceed the period $P$ of the periodic event handler.

\begin{figure}[!tb]
	\begin{circus}
		\circsafelet\ Safelet \circdef \ldots\\ 
		\circsequencer\ Sequencer \circdef \ldots\\ 
		\circmission\ Mission \circdef \ldots\\ 
		\circperiodic\ PeriodicHandler \circdef \circbegin\\\quad
		\circstart~0~\circperiod ~P\\\quad
		\circstate\ [ah: ID]\\\quad
		\circinitial \circdef ah: ID @ this.ah := ah\\\quad
		\circhandleAsyncEvent \circdef\\\quad\quad
		((input?x \then \Skip)\circstartby ID\circseq setBuffer!(buffer\cat\langle x\rangle) \then release() \circseq (\circwait 0\upto PTB))
		\circendby\ PD\\
		\circend\\
		\circaperiodic\ AperiodicHandler \circdef \circbegin\\\quad
		\circhandleAsyncEvent \circdef\\\quad\quad
		\circblockbegin
		getBuffer?buffer \then\\
		\circblockbegin\circif buffer \in theSame \circthen (output!true \then \Skip)\circstartby OD\\
		\circelse buffer \notin theSame \circthen (output!false \then \Skip)\circstartby OD\\
		\circfi\circblockend\circseq
		\circwait 0..ATB
		\circblockend
		\circendby\ AD\\
		\circend
	\end{circus}\vspace{-1cm}
	\caption{SCJ Level 1 example: S-anchor}
	\label{fig:S-anchor}
\end{figure}

\begin{figure}[!t]
\begin{syntax}
	\scat{SCJProgram} \sdef \scat{SCJParagraph}^*\\
	\scat{SCJParagraph} \sdef \scat{Safelet} | \scat{MissionSequencer}
	| \scat {Mission} | \scat{Handler} | \scat{CircusParagraph}
\end{syntax}\vspace{-0.8cm}
\begin{syntax}
	\scat{Safelet} \sdef \circsafelet\ \scat{N} \circdef \circbegin\\
	&& \quad \scat{SCJSSafeletProcessParagraph}^*
	\\ %
	&& \quad \circstate~\scat{Schema\hbox{-}Expression}
	\\ %
	&& \quad \scat{SCJSafeletProcessParagraph}^*
	\\ %
	&& \quad \circinitialize \circdef \scat{SCJSafeletAction}
	\\ %
	&& \quad \scat{SCJSafeletProcessParagraph}^*
	\\ %
	&& \quad \circgetSequencer \circdef \circres\ \circreturn: \circsequencerref @ \scat{SCJSafeletAction}
	\\ %
	&& \quad \scat{SCJSafeletProcessParagraph}^*
	\\ %
	&& \circend
\end{syntax}
\label{fig:syntax}\vspace{-.5cm}
\caption{Syntax of \SCJCircus\ (sketch)}
\end{figure}

In general, as shown in Figure~\ref{fig:syntax}, an \SCJCircus\ program is a sequence of $\scat{SCJParagraphs}$, which can be a \Circus\ paragraph, or the declaration of a safelet, mission sequencer, mission or handler. The structure of each of the SCJ-specific abstractions is determined by the values and behaviours that must be specified for an application according to the SCJ standard~\cite{SCJ}. For instance, a safelet must implement the \texttt{initialize} method that allows the allocation of global objects, and the \texttt{getSequencer} method that provides a mission sequencer.

Accordingly, the \SCJCircus~construct corresponding to a safelet in Figure~\ref{fig:syntax} has a name taken from the set of valid \Circus\ names $\scat{N}$, and allows the specification of state components ($\circstate$), the initialisation ($\circinitialize$) and $\circgetSequencer$ methods, as well as auxiliary actions ($\scat{SCJSafeletProcessParagraph}$). The state components model the fields of the safelet class. An $\scat{SCJSafeletProcessParagraph}$ allows the specification of an action whose body is a $\scat{SCJSafeletAction}$, which restricts the constructs that can be used in an action of a safelet, in particular, the type of allocation constructs as discussed next.

SCJ enforces a hierarchical memory-model in which different components (safelets, missions and so on) may only instantiate new objects in their memory areas or parent memory areas. We reflect this discipline in \SCJCircus\ by restricting syntactically which paragraphs may include allocations, through different $\circnew$ keywords, to particular memory areas. For instance, a safelet may only instantiate objects in the immortal memory, and therefore may only use the keyword $\circnewI$ for instantiation of new objects. A handler, on the other hand, may allocate objects in the immortal memory area, mission memory area ($\circnewM$), per-release memory area ($\circnewPR$) and private memory area~($\circnewPM$). 

These restrictions are reflected in the use of different syntactic categories for the actions and paragraphs of the different constructs. For example, the \texttt{getSequencer} method of a safelet must be an $\scat{SCJSafeletAction}$ and the \texttt{handleAsyncEvent} method of a handler must be as $\scat{SCJHandlerAction}$. The first only allows instantiation via $\circnewI$, whilst the other allows all possible instantiation keywords.

The syntax of the \SCJCircus\ paragraphs for the mission sequencer, missions and handlers are similar, providing means for the specification of state components ($\circstate$), constructors ($\circinitial$), and the methods of the corresponding element that must be provided by the developer. For further details about the syntax of \SCJCircus\ refer to \cite{MiyazawaReport2015}.

\subsection{Semantic model}
\label{sec:model}

In \cite{ZLCW13}, an approach to modelling SCJ programs has been proposed; it is a translation strategy defined as a semantic function that maps SCJ programs to \Circus\ specifications.  We adopt a similar approach here to give semantics to \SCJCircus.  Our \Circus\ models, however, are updated to consider recent significant changes to SCJ and to cater for compositional reasoning about SCJ constructs. 

Each \CircusTime\ process is defined as the parallel composition of two processes: a framework process that captures the behaviour of the corresponding SCJ component as an element of the SCJ programming model, and a process that captures the behaviour of that component as defined in a particular application. For example, the process $PEHFW$ in Figure~\ref{fig:PEHFW} presents the framework process for a periodic handler. It defines the general flow of execution of such a handler without giving the details of a particular handler implementation.  

\begin{figure}\centering
	\includegraphics[width=0.9\textwidth]{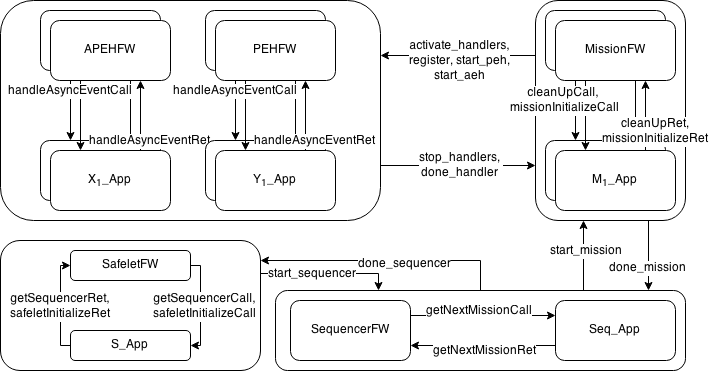}
	\caption{Structure of semantic models}
	\label{fig:structure}
\end{figure}

The framework and application processes of each SCJ element interact through a number of channels that correspond to method calls in the implementation. For example, the channels $safeletInitializeCall$, $safeletInitializeRet$, $getSequencerCall$ and $getSequencerRet$ in Figure~\ref{fig:structure} are used by the safelet framework process $SafeletFW$ to communicate with the application specific process $S\_App$ and correspond to calls to the methods \texttt{initialize} and \texttt{getSequencer} of the application.

In the models of SCJ programs presented in \cite{ZLCW13}, the application processes are combined together in interleaving, framework processes are grouped together in parallel, and both groups are then combined in parallel to yield the semantic model of the whole application. This structure proved not ideal for the compositional analysis of \SCJCircus\ programs because the aspects relevant to a specific \SCJCircus\ construct, such as a handler, are spread through the complete model and cannot be isolated for reasoning purposes. Figure~\ref{fig:structure} depicts the structure of the updated semantic model. \Circus\ specifications model each of the \SCJCircus\ paragraphs as standard \CircusTime\ processes. 

It is worth mentioning that the model where application and framework processes are composed on a per-element basis is a refinement of the model structured as in \cite{ZLCW13}. This fact is established in our refinement strategy described in Section~\ref{sec:refinement} because the framework is first introduced as a monolithic process, and then distributed through the application processes.

\begin{figure}[!t]
\begin{circus}
	\circprocess SafeletFW \circdef id: ID @ \circbegin\\
	\t1 Execute \circdef getSequencerCall!id!id \then getSequencerRet!id!id?s \then\\
	\t2 \circblockbegin
	\circif s \neq null \longrightarrow start\_sequencer \then done\_sequencer \then \Skip\\
	\circelse s = null \longrightarrow \Skip\\
	\circfi
	\circblockend\\
	\t1 @ safeletInitializeCall!id!id \then safeletInitializeRet!id!id \then Execute;\\
	\t2 end\_safelet\_app \then \Skip\\
	\circend
\end{circus}\vspace{-1cm}
\caption{Framework process for Safelet.}
\label{fig:safeletfw}
\end{figure}

The framework process that specifies the generic behaviour of a safelet is shown in Figure~\ref{fig:safeletfw}. It is a process parametrised by the safelet identifier, and its behaviour consists of requesting to the application process the initialisation of the safelet using the channels $safeletInitializeCall$ and $safeletInitializeRet$, obtaining a mission sequencer via the channels $getSequencerCall$ and $getSequencerRet$, and if the sequencer is different than $null$, starting it (using the channel $start\_sequencer$). At this point the safelet framework process waits for the mission sequencer to complete its execution and signal on the channel $done\_sequencer$, in which case the safelet indicates to the application process that it is terminating through the channel $end\_safelet\_app$, and terminates ($\Skip$).

The complete definition of the model can be found in \cite{MiyazawaReport2015}.

\subsection{Semantics}
\label{sec:semantics}

The semantics of \SCJCircus\ is formalised as a function from well-formed models written in accordance with the abstract syntax of \SCJCircus\ to \Circus\ models, that is, elements of the category $CircusProgram$, as defined in \cite{Oli06}. In order to improve readability, the semantics is presented in terms of translation rules that output \Circus\ concrete syntax. In essence, the semantic function composes the behaviours specified in \SCJCircus\ with the model of the SCJ framework discussed in Section~\ref{sec:model} in a compositional way.

Formally, the semantics of an \SCJCircus~program $p$ is given by the \Circus\ program formed by the \Circus\ paragraphs that are obtained by applying specific semantic functions to the paragraphs of $p$. This is specified below by the function
$\llbracket\varg\rrbracket_{SCJProgram}$ that takes a well-formed \SCJCircus\ program and outputs a \Circus\ program composed of the paragraphs produced by the semantic functions $\llbracket\varg\rrbracket_{SCJParagraphs}$ and $\llbracket\varg\rrbracket_{Application}$. The first takes a sequence of \SCJCircus\ paragraphs and outputs a sequence of \Circus\ paragraphs, and the second takes a program and outputs the definition of a process that composes the processes defined in the previous paragraphs to specify the overall meaning of the application.
\begin{axdef}
	\llbracket\varg\rrbracket_{SCJProgram}: SCJProgram \pfun Program\\
	\where
	\forall p: WF\_SCJProgram @ \llbracket p \rrbracket_{SCJProgram} = \llbracket p.paragraphs \rrbracket_{SCJParagraphs}\cat\llbracket p\rrbracket_{Application}\\
\end{axdef}
We use the mathematical notation of Z \cite{WD96} to specify our semantic functions, and explain any non-standard use of notation as needed. In what follows, we focus on the semantic function for the safelet, which is used by $\llbracket \varg \rrbracket_{SCJParagraphs}$ to give semantics to a safelet paragraph.

As explained in the previous section, the semantics of a safelet is given by the parallel composition of a \Circus\ process that characterises the application-specific behaviours and a \Circus\ process that models the generic behaviour of the framework. It is given by the function $\llbracket\varg\rrbracket_{Safelet}$ below, which takes a safelet $s$ and outputs a sequence of two processes: the application process $s\_app$ and the process that models the complete behaviour of $s$ as the parallel composition of $s\_app$ and the framework process $SafeletFW$ instantiated by the identifier of $s$. In the definition of the semantic function, guillemots ($\meta{~}$) are used to distinguish the \Circus\ syntax from the meta-language used to specify the rules. For instance, 
$\meta{safelet\_app(s)}$, indicates that the function $safelet\_app$ must be evaluated on the parameter $s$ and the resulting syntax tree must be substituted in place of $\meta{ safelet\_app(s)}$.

\begin{axdef}
	\llbracket\varg\rrbracket_{Safelet}: Safelet \pfun \seq CircusParagraph\\
	\where\\
	\t1\forall s: WF\_Safelet @\\
	\llbracket s \rrbracket_{Safelet} =
	\circblockbegin
	\circprocess ~ \meta{ name(s)}\_App ~ \circdef ~ \meta{ safelet\_app(s)}\\
	\circprocess ~ \meta{ name(s)} ~ \circdef\\
	\t1 (SafeletFW(\meta{name(s)}ID) \lpar \meta{ SafeletCS(s)} \rpar \meta{ name(s)}\_App)\\\t2 \circhide \meta{ SafeletCS(s)}
	\circblockend
\end{axdef}
As shown above, the definition of $\llbracket\varg\rrbracket_{Safelet}$ relies on the function $safelet\_app$ that produces the application specific process, and a function $SafeletCS$ that calculates the channels on which the application and the framework must communicate. These channels are internal to the safelet and therefore hidden ($\circhide$).

\begin{figure}[!t]
\begin{axdef}
	safelet\_app: Safelet \pfun BasicProcess
	\where
	\forall s: WF\_Safelet @\\
	safelet\_app(s) = \\
	\circblockbegin\circbegin\\
	\quad \circstate~\meta{s.state}\\
	\quad \meta{ for~each~p: s.paragraphs~of~(N \circdef SCJSafeletParametrisedAction)~do}\\
	\quad\quad \meta{N}Meth \circdef \meta{translate\_method(name(s)ID,N,p.body)}\\
	\quad\meta{ end}\\
	
	\quad getSequencerMeth \circdef \meta{translate\_method(name(s)ID,getSequencer,s.getSequencer)}\\
	
	\quad initializeApplicationMeth \circdef initializeApplicationCall?x!\meta{name(s)}ID \then \\\t2
	\meta{s.initialize}; initializeApplicationRet!x!\meta{name(s)}ID \then \Skip\\
	
	\quad Methods \circdef \mu X @\\
	\quad\quad getSequencerMeth; X \extchoice~initializeApplicationMeth; X\\
	\quad \meta{ for~each~p: s.paragraphs~of~(N \circdef A)~do}
    \extchoice \meta{N}Meth; X
	\meta{end}\\
	\quad\quad \extchoice
	end\_safelet\_app \then \Skip\\
	\quad @ Methods\\	
	\circend
	\circblockend
\end{axdef}\vspace{-.5cm}
\caption{Semantic function $safelet\_app$.}
\label{fig:safelet-app}
\end{figure}

The $safelet\_app$ function shown in Figure~\ref{fig:safelet-app} takes a safelet $s$ and constructs a process named after $s$ using the function $name$ concatenated with $\_App$, and with the same state as $s$. Each auxiliary method of the safelet is translated into an \Circus\ action using a pair of channels to model the call and return of the method. Similarly, the methods \texttt{getSequencer} and \texttt{initialize} are translated into the actions $getSequencerMeth$ and $initializeApplicationMeth$. All these actions are used to construct the action $Methods$ that recursively offers a choice between each of those actions, and the possibility to terminate the recursion via a synchronisation on the channel $end\_safelet\_app$.

The overall behaviour of the process is the action $Methods$. The parallel composition of the process obtained from the safelet and the framework process synchronises on the call and return channels used to encode method calling, as well as on the channel $end\_safelet\_app$, and these channels are made internal using the hiding operator ($\circhide$).

The functions $sequencer\_app$, $mission\_app$, $PEH\_app$ and $AEH\_app$ that define the application processes for mission sequencers, mission, periodic event handler and aperiodic event handlers are defined similarly and are omitted. The complete semantics is defined in~\cite{MiyazawaReport2015}.

\section{Refinement Strategy}
\label{sec:refinement}

The refinement strategy proposed in \cite{Cavalcanti2013} covers the refinement of abstract \CircusTime\ models into a process written following a pattern in which some of the structure of an SCJ application is identified but not explicitly described in terms of independent SCJ components, as can be done using \SCJCircus. Here, we further elaborate the original strategy to obtain an S-Anchor like that shown in Figure~\ref{fig:S-anchor}.

Our refinement strategy starts from an E-Anchor in the form shown in Figure~\ref{fig:pattern-E-anchor}, which is a single \Circus\ process in which each action models a component of the desired SCJ implementation, but the different elements (e.g., safelet, mission sequencer, and so on) are not yet isolated. The only parallelism is between the two handlers. The E-Anchor of our running example obtained through the application of the refinement strategy in \cite{Cavalcanti2013} to the abstract model is shown in Figure~\ref{fig:E-anchor}. It is a single \Circus\ process whose main action calls the safelet, which then calls the mission sequencer. The mission sequencer calls the single mission of our example, which calls in parallel the periodic and aperiodic handlers as well as an action that models the mission memory shared by both handlers.

In order to obtain the S-anchor, we propose a refinement strategy based on four phases: (1) introduction of SCJ control flow, (2) introduction of application process, (3) introduction of framework processes, (4) conversion to \SCJCircus. The resulting S-anchor for our example is shown in Figure~\ref{fig:S-anchor}.

The first phase introduces the patterns of control observed in \SCJCircus\ models, such as call-return channels, which model method calls, $start$ and $done$ channels that model the execution and termination of \SCJCircus\ abstractions (e.g., Safelet), and release mechanisms. The second phase separates application-specific behaviours (e.g., reading of input) from framework behaviours (e.g., request of mission sequencer in the safelet). The third phase takes the incomplete model of framework behaviour isolated in the second phase and expands it by completing them with all possible behaviours of the \SCJCircus\ framework processes. This is necessary because the E-anchor does not cover aspects of the framework that are not used by the application. For instance, our running example does not model termination and, therefore, the framework-specific behaviour isolated in the second phase does not cover the termination mechanisms of the SCJ framework. These are introduced in the third phase. The fourth phase introduces the paragraphs of the S-anchor, where the \SCJCircus\ abstractions are explicitly declared.

\subsection{E-anchor: starting point}

We identify four main patterns of E-anchors with respect to the synchronisation between a number of periodic and aperiodic event handlers. The first has both types of handlers executing cyclically in lockstep and terminating within the period of the periodic event handler. In this pattern, all handlers are executed at every cycle and must terminate before the next cycle.

The second pattern is similar, except that not all aperiodic handlers are executed at each cycle. The handlers are executed cyclically, but not in lockstep. The common property to the first two patterns is that the execution of both periodic and aperiodic event handlers finishes with the period of the application. The third and fourth patterns are version of the first two where the deadline of the aperiodic event handlers cannot be guaranteed. That is, the execution of an aperiodic event handler may not terminate before the next cycle starts. In this paper, we focus on E-anchor of the first type: cyclic in lockstep. The model in Figure~\ref{fig:S-anchor} follows this pattern. Examples of the remaining patterns can be found in \cite{MiyazawaReport2015}.

\begin{figure}[!tb]\vspace{-.5cm}\centering
	\begin{minipage}{.5\textwidth}
	\begin{circus}
		\circprocess P \circdef \circbegin\\\quad
		\circstate S\\\quad
		Handler_i \circdef \ldots\\\quad
		MArea_j \circdef \ldots\\\quad
		Mission_j \circdef (MArea_j \parallel (\parallel k: handlers_j @ Handler_{k}))\\\quad
		
		MissionSequencer \circdef \Semi i: 1\upto n @ Mission_i\\\quad
		Safelet \circdef MissionSequencer\\\quad
		Application \circdef Safelet\\\quad
		\circspot Application\\
		\circend
	\end{circus}
	\end{minipage}
	\caption{Refinement strategy: starting point of first phase (E-anchor)}
	\label{fig:pattern-E-anchor}
\end{figure}

\begin{figure}[!tb]	
	\begin{circus}
		\circprocess System1 \circdef \circbegin\\\quad
		MArea \circdef \circvar buffer: \seq \nat @ \circmu X @\\\quad\quad
		(setBuffer?x \then buffer := x\circseq X
		\extchoice
		getBuffer!buffer \then X
		\extchoice
		stop \then \Skip)\\\quad
		
		PeriodicHandler \circdef\\\quad\quad
		\circmu X @ \circblockbegin
		\circblockbegin
		(input?x \then \Skip)\circstartby ID\circseq\\
		setBuffer!(buffer\cat\langle x\rangle) \then 
		release \then (\circwait 0..PTB)
		\circblockend\circendby PD\\
		\interleave
		\circwait P
		\circblockend\circseq X\\
		\quad
		
		AperiodicHandler \circdef\\\quad\quad
		\circmu X @ 
		\circblockbegin 
		release \then \\\quad
		\circblockbegin getBuffer?buffer \then \circwait 0..ATB\circseq\\ 
		\circblockbegin\circif buffer \in three_0 \circthen (output!true \then \Skip)\circstartby OD\\
		\circelse buffer \notin three_0 \circthen (output!false \then \Skip)\circstartby OD\\
		\circfi\circblockend\circblockend\circendby AD\circseq X\\
		\circblockend\\\quad
		
		
		
		Mission = 
		\circblockbegin
		\circblockbegin
		PeriodicHandler\\
		\lpar \{\} | \lchanset stop, release \rchanset | \{\} \rpar\\ AperiodicHandler
		\circblockend \circhide \lchanset release \rchanset\\
		\lpar \{\} | \lchanset \ldots\rchanset | \{\} \rpar
		MArea\circblockend \circhide \lchanset setBuffer, getBuffer\rchanset
		\\\quad
		MissionSequencer \circdef Mission\\\quad
		Safelet \circdef MissionSequencer\\\quad
		Application \circdef Safelet\\\quad
		\circspot Application\\
		\circend
	\end{circus}\vspace{-1cm}
	\caption{SCJ Level 1 example: E-anchor}
	\label{fig:E-anchor}
\end{figure}

\subsection{(CF) Introduction of SCJ control flow}

This phase introduces some of the parallel structure observed in \SCJCircus\ programs. Figure~\ref{fig:phase-one-target} shows the structure of the process obtained by applying the first phase. 

We recall that, as the first phase of the refinement, its starting point is an E-anchor described in Figure~\ref{fig:pattern-E-anchor}, and illustrated in Figure~\ref{fig:E-anchor} for our example. The target is shown in Figure~\ref{fig:phase-one-target}. This is a model is still a single process, but its main action composes a number of auxiliary actions in parallel, each of which specifies the behaviours of an SCJ abstraction.

In this phase, parallelism introduction laws such as Law~1 are used to refine an action $F(A)$ into a parallelism where the subaction $A$ is replaced by two communications on channels $c_1$ and $c_2$, and the parallel action is formed by the first communication on $c_1$, followed by the subaction $A$, followed by the second communication on $c_2$.

\noindent\begin{minipage}{\textwidth}
	\vspace{.5cm}
	\paragraph{Law 1. Parallelism Introduction.\vspace{-.3cm}}
	\begin{circus}\hspace{-.6cm}
		F(A) \circrefines 
		(F(c_1 \then c_2 \then \Skip) \lpar usedV(F) | \lchanset c_1, c_2\rchanset | userV(A) \rpar c_1 \then A\circseq c_2 \then \Skip)\circhide \lchanset c_1, c_2\rchanset
	\end{circus}
	\paragraph{provided}
		$usedV(F) \cap usedV(A) = \emptyset \land \lchanset c_1, c_2\rchanset \cap usedC(F(A)) = \emptyset$
	\vspace{.3cm}
\end{minipage}

\noindent This law can be proved by structural induction over the structure of the action $F$ using distribution and step laws such as the ones found in \cite{Oli06}. The provisos guarantee that $c_1$ and $c_2$ are not used in $A$, and that the variables used in the action $F$ and the subaction $A$ form a partition of the state, so that they can be put in parallel without creating race conditions. As shown, the \Circus\ parallel operator for actions defines partitions of the state for use of each of the parallel actions.

\begin{figure}[!t]
	\begin{circus}
		\circprocess CF\_P \circdef \circbegin\\\quad
		\circstate S\\\quad
		CF\_Safelet \circdef getSequencerCall \then 
			geSequencerRet?x \then start\_sequencer!x \then\\\quad\quad
			done\_sequencer!x \then \Skip\\\quad
		CF\_MissionSequencer \circdef \ldots\\\quad
		CF\_Mission_j \circdef \ldots\\\quad
		CF\_Handler_i \circdef \ldots\\\quad
		CF\_Application \circdef\circblockbegin
		CF\_Safelet
		\parallel
		CF\_MissionSequencer\\
		\parallel
		((\parallel i: 1\upto n @ CF\_Mission_i)
		\parallel
		(\interleave i: 1 \upto m @ CF\_handler_i))
		\circblockend\\\quad
		\circspot CF\_Application\\
		\circend
	\end{circus}\vspace{-1cm}
	\caption{Refinement strategy: target of CF phase}
	\label{fig:phase-one-target}
\end{figure}  

In general, Law~1 must be applied to the actions that model the safelet, the mission sequencer, the missions, and the handlers. In our example, this law is applied to the action $Safelet$ in Figure~\ref{fig:E-anchor} to separate it from $MissionSequencer$, and then to the action $MissionSequencer$ to separate it from $Mission$, and finally to the action $Mission$ to separate it from $Handlers$. At this point, we obtain the action $CF\_Application$. The resulting structure is depicted in Figure~\ref{fig:phase-one-target}, where the actions prefixed by $CF\_$ are the actions in Figure~\ref{fig:E-anchor} modified by the application of the refinement laws.


\subsection{(AP) Introduction of application processes}

The target of this phase is shown in Figure~\ref{fig:AP-target}: it defines a number of application processes, and refines the process $CF\_P$ into the parallel composition of the interleaved application processes and a modified version of $CF\_P$ ($CF\_P\_FW$), where application-specific behaviours have been replaced by calls to actions of the application processes via channel communications using $Call$ and $Ret$ channels.

\begin{figure}[!t]
	\begin{circus}
		\circprocess\ Handler_i\_app \circdef \ldots\\
		\circprocess\ Mission_j\_app \circdef \ldots\\
		\circprocess\ MissionSequencer\_app \circdef \ldots\\
		\circprocess\ Safelet\_app \circdef \ldots\\
		\circprocess\ AP\_P \circdef CF\_P\_FW \parallel
		\circblockbegin
		Safelet\_app~\interleave MissionSequencer\_app~\interleave\\
		(\interleave i: 1\upto m @ Handler_i\_app)
		\interleave
		(\interleave i: 1\upto n @ Mission_i\_app)
		\circblockend
	\end{circus}\vspace{-0.5cm}
	\caption{Refinement strategy: target of phase \textbf{AP}}
	\label{fig:AP-target}
\end{figure}

In this phase, we use the process obtained in phase \textbf{CF} to identify the behaviours that are application specific and construct application processes. Next, each application process is introduced in parallel with the original process and the behaviour provided by the application process is replaced in the original process by calls via the appropriate channels. This is achieved using refinement laws similar to Law \texttt{server-intro} in \cite{Miyazawa2012}, which supports the introduction of a server-client architecture.

\begin{figure}
\begin{circus}
CF\_System\left[\begin{array}{l}\circblockbegin
(input?x \then \Skip) \circstartby ID \circseq\\
setBuffer!(buffer\cat\langle x\rangle) \then release \then (\circwait 0 \upto PTB)
\circblockend\circendby PD\end{array}\right]\\
\circrefines\\
\circblockbegin
CF\_System\left[\begin{array}{l}handleAsyncEventCall?x!PHID \then\\ handleAsyncEventRet!x!PHID \then \Skip\end{array}\right]\\
\lpar\{\}|\lchanset handleAsyncEventCall, handleAsyncEventRet | \{\}\rpar\\
PeriodicHandler\_App
\circblockend
\circhide\lchanset \ldots \rchanset
\end{circus}\vspace{-.5cm}
\caption{Introduction of application process for $PeriodicHandler$ in our example.}
\label{fig:application-intro}
\end{figure}


Figure~\ref{fig:application-intro} illustrates the application of this phase to the $CF\_PeriodicHandler$ action obtained after the application of the first phase to the example in Figure~\ref{fig:E-anchor}. First, the process $PeriodicHandler\_App$ is generated as a recursion that, at each step, offers the event $handleAsyncEventCall$, executes the behaviour of the original periodic handler, and synchronises on $handleAsyncEventRet$. Next, the process $CF\_System$ obtained by the first phase containing the behaviour of the periodic handler (made explicit in Figure~\ref{fig:application-intro} by the square brackets after $CF\_System$) is refined into the parallel composition of the generated application process $PeriodicHandler\_App$ and $CF\_System$ with the behaviour of the periodic event handler in brackets replaced by the synchronisations on $handleAsyncEventCall$ and $handleAsyncEventRet$ ($PHID$ is the identifier of the periodic handler and is necessary to support multiple handlers).

\subsection{(FW) Introduction of framework processes}

\begin{figure}[!t]
	\begin{circus}
		\circprocess\ FW\_P \circdef\circblockbegin
		\circblockbegin
		SafeletFW \parallel SequencerFW \parallel
		(\interleave i: 1\upto n @ MissionFW(mission_i)) \parallel\\
		(\interleave i: 1\upto m @ HandlerFW(handler_i))
		\circblockend\\
		\parallel
		\circblockbegin
		Safelet\_app~\interleave MissionSequencer\_app~\interleave\\
		(\interleave i: 1\upto m @ Handler_i\_app)
		\interleave
		(\interleave i: 1\upto n @ Mission_i\_app)
		\circblockend\circblockend
	\end{circus}\vspace{-.5cm}
	\caption{Refinement strategy: target of phase \textbf{FW}}
	\label{fig:FW-target}
\end{figure}

The target of this phase is shown in Figure~\ref{fig:FW-target}; it consists of the interleaved application processes in parallel with the parallel composition of the framework processes discussed in Section~\ref{sec:model}.

This phase acts on the process of $CF\_P\_FW$ in Figure~\ref{fig:AP-target}, from which all application-specific behaviours have been removed (and distributed to the application processes). What remains in $CF\_P\_FW$ after the second phase are the framework behaviours that are relevant to the particular application. In this phase, we complement these framework behaviours to account for the behaviours that are part of the framework, but not used in $CF\_P\_FW$. This is achieved by the application of refinement laws such as Law~2 to introduce the actions that correspond to the control flow present in the framework processes but not used by the application processes. 

\noindent\begin{minipage}{\textwidth}
	\vspace{.5cm}
	\paragraph{Law 2. Unused behaviour introduction.\vspace{-.3cm}}
	\begin{circus}\hspace{-.6cm}
		(a \then A \lpar ns_1 | cs | ns_2 \rpar a \then B) \circrefines 
		(a \then A \lpar ns_1 | cs \cup \lchanset b\rchanset | ns_2 \rpar (a \then B \extchoice b \then C)) 
	\end{circus}
	\paragraph{provided} $a \in cs \land b \notin usedC(A,B)$
	\vspace{.3cm}
\end{minipage}

\noindent Law~2 allows the introduction of actions in a parallelism that are never used; it relies on the fact that the channel $b$ is not used anywhere else in the left hand side. Since $b$ is in the synchronisation set of the refined action, the action $b \then C$ can never be started. 


Finally, process parallelism introduction laws are used to refine the process $CF\_P$ into the process $FW\_P$ defined as a parallel composition of processes whose main actions are $FW\_A$. The structure of the refined process follows the structure of the main action of $CF\_P$. Figure~\ref{fig:FW-target} shows the structure of process $FW\_P$ obtained in this phase, where $mission_i$ is the identifier of the i-th mission, $handler_i$ is the identifier of the i-th handler, and $HandlerFW$ is either $PEHFW$ or $APEHFW$ depending on whether the i-th handler is periodic or aperiodic.

\subsection{(Conv) Conversion to \SCJCircus.}

\begin{figure}[!t]
	\begin{circus}
		\circhandler\ S\_Handler_i\circdef \ldots\\
		\circmission\ S\_Mission_j \circdef \ldots\\
		\circsequencer\ S\_MissionSequencer \circdef \ldots\\
		\circsafelet\ S\_Safelet \circdef \ldots
	\end{circus}\vspace{-1cm}
	\caption{Refinement strategy: target of phase \textbf{Conv}}
	\label{fig:Conv-target}
\end{figure}

The target of this phase is shown in Figure~\ref{fig:Conv-target}; it explicitly refers to the SCJ abstractions that have been incorporated in \SCJCircus. In this final phase of our refinement strategy, the top-level parallel actions of $FW\_P$ in Figure~\ref{fig:FW-target} are merged into a parallel composition of pairs of application and framework processes. This is achieved by the application of a procedure similar to the one used in a refinement strategy described in~\cite{Miyazawa2013}. It relies on the syntactic structure of the parallel actions and the use of refinement laws to eliminate or distribute the parallel composition over other \Circus\ constructs such as external choice, recursion and interleaving.

Next, each parallel composition of application and framework processes ($A\_app \parallel A\_FW$) is used to define a new process $A$, and the process $FW\_P$ is refined by replacing the parallelisms of the form $A\_app \parallel A\_FW$ by a call to the newly defined processes. The resulting processes are shown in Figure~\ref{fig:par-elim}. At this point, each SCJ abstraction is defined by its own process that includes the application and framework-specific behaviours in different parallel processes.

\begin{figure}[!t]
	\begin{circus}
		\circprocess\ Safelet \circdef SafeletFW \parallel Safelet\_app\\
		\circprocess\ MissionSequencer \circdef SequencerFW \parallel MissionSequencer\_app\\
		\circprocess\ Mission_j \circdef MissionFW(mission_j) \parallel Mission_j\_app\\
		\circprocess\ Handler_i \circdef HandlerFW(handler_i) \parallel Handler_i\_app\\
		\circprocess\ FW\_P \circdef (
		Safelet \parallel Sequencer \parallel
		(\interleave i: 1\upto n @ Mission_i) \parallel
		(\interleave i: 1\upto m @ Handler_i))\\
	\end{circus}\vspace{-1cm}
	\caption{Refinement strategy: parallelism elimination in phase \textbf{FW}}
	\label{fig:par-elim}
\end{figure}

Finally, the semantics of \SCJCircus\ is used to refine each newly defined process $A$ into the corresponding \SCJCircus\ abstraction, and the sequence of \SCJCircus\ abstractions into a complete \SCJCircus\ program. Figure~\ref{fig:Conv-target} shows the general structure of the program resulting from this phase, and Figure~\ref{fig:S-anchor} shows the result of applying our refinement strategy to the E-Anchor of our running example.

\section{Conclusions}
\label{sec:conclusions}

In this paper, we extend previous work~\cite{ZLCW13,Cavalcanti2013} on both the semantics of SCJ and refinement strategies for SCJ programs. We propose a variant of \Circus\ suitable for modelling SCJ concepts, update existing models of SCJ to reflect changes to the SCJ specification and better suit the goal of compositional verification, formalise the semantics of \SCJCircus\ in terms of these updated models, and extend a previously proposed refinement strategy to account for the refinement to \SCJCircus\ specifications. 

Other significant differences between our model of SCJ and that in \cite{ZLCW13} include: (1) the shift from the use of events to trigger the execution of aperiodic event handlers in previous version of the SCJ specification, to the direct use of the asynchronous method \texttt{release} of the aperiodic event handler, and (2) modelling of handlers using two processes $PEHFW$ (periodic event handler) and $APEHFW$ (aperiodic event handler) so that the distinction between periodic and aperiodic event handlers are made at the framework level, instead of the application level.

The SCJ standard specifies the new constructs, the API, and the SCJ VM, but says nothing about verification and design of programs. Our effort complements those in
\cite{Kalibera,Tang,Haddad,MC14a}. Kalibera \emph{et al.}~\cite{Kalibera} apply model checking and exhaustive testing to perform scheduling and race-condition analysis in SCJ programs. Haddad \emph{et al.}~\cite{Haddad} extend the Java Modeling Language~\cite{JML} with timing properties to support worst-case execution analysis of SCJ programs, whilst Tang \emph{et al.}~\cite{Tang} use annotations to analyse SCJ programs for memory safety and compliance to SCJ levels. Marriott \emph{et al.}~\cite{MC14a}, on the other hand, perform automatic verification of memory-safety without requiring the user to annotate the program.

We identify four main application patterns with respect to the timing properties of the periodic and aperiodic event handlers. Whilst we focus our effort here on the refinement of one particular pattern (cyclic in lockstep), the refinement strategy is general enough to be applied to the other patterns with localized changes. We will address this issue in the context of SCJ in future work. We will also detail the refinement strategy and mechanise it in a theorem prover in order to further validate it.

\paragraph{Acklowledgements.} This work is funded by the EPSRC grant EP/H017461/1. No new primary data was created during this study.

\nocite{*}
\bibliographystyle{eptcs}

\bibliography{refine}
\end{document}